# Quantum Disentanglement Eraser: an Optical Implementation Using Three-photon Spontaneous Decay.

Alejandro A. Hnilo.

*CEILAP, Centro de Investigaciones en Láseres y Aplicaciones, UNIDEF (MINDEF-CONICET); CITEDEF, J.B. de La Salle 4397, (1603) Villa Martelli, Argentina.*
emails: ahnilo@citedef.gob.ar, alex.hnilo@gmail.com
October 26th, 2014.

The "disentanglement eraser" or "entanglement restorer" scheme allows retrieving entanglement by erasing the information about the formation of a classical (or separable) state. It implies a close analogy between the pairs of properties: entangled-separable and wave-particle, the latter as it is described in the well known "quantum eraser" scheme. Both schemes illustrate the physical importance of information in quantum systems. The disentanglement eraser is also useful as a building block in quantum information and computation. Its full optical implementation using three-photons decay in a centro-symmetrical crystal is proposed. It seems technically simpler to achieve than the known proposed alternatives.

Index Topics: Quantum Optics, Nonlinear Optics, Quantum Information.
Keywords: Optical implementation of Quantum computation and Information, Optical tests of Quantum Mechanics, Optical experiments on Bell's inequalities and GHZ states.

## 1. Introduction.

One of the intriguing features of Quantum Mechanics is the wave-particle dualism. The well-known "quantum eraser" scheme illustrates how a quantum system in a Mach-Zehnder interferometer behaves as a particle or as a wave depending on the available *information* about the system. The quantum eraser has been beautifully demonstrated in an optical experiment [1]. Other of the intriguing features of Quantum Mechanics is entanglement [2]. Garisto and Hardy [3] proposed a scheme named "disentaglement eraser" (also: "entanglement restorer"). In this scheme, the interference term (in this case, between the two possible states of a quantum state of *two* particles) can be "turned on" or "off" depending whether an appropriate measurement on an auxiliary system is performed or not, in a way totally analogous to the quantum eraser for the two paths of a single particle in an interferometer. That interference term is intrinsic to entanglement, so that the entanglement of the state of the two particles can be turned on or off. The scheme not only illustrates the decisive role played by the information in the observable properties of quantum systems (in this case, the degree of entanglement), but it is also useful in quantum computation and quantum information processing in general. In fact, the scheme's early description was based on C-NOT gates. The scheme was also shown of interest to generate cluster and graph states [4].

The earliest proposal for the disentanglement eraser [3] involved a pair of photons produced by two-photon spontaneous decay (2psd) in a non-centro-symmetrical crystal. The (entangled) degrees of freedom of position and polarization were proposed to achieve both reversible and irreversible disentanglement. Yet, the GHZ state produced in this way would be observable in two positions instead of three, and hence (as the authors themselves point out) it would not be fully satisfactory. The scheme was later achieved in a liquid-state NMR experiment [5]. Yet, it was done for equivalent ensemble states, not for true single system states. The use of single atom high-Q microwave cavities was later proposed to realize the scheme [4,6]. However, this proposal seems to be more difficult to perform and to scale up than a full optical alternative.

In this paper, I propose a full optical implementation of the disentanglement eraser by using a single process of *three* photon spontaneous decay. It appears technically simpler than the known alternatives. In the next Section 2, the idea of the disentanglement eraser is reviewed. The proposal is described in the Section 3. Some of the distinctive features of this proposal are discussed in the Section 4.

## 2. The disentanglement eraser.

Consider for example the fully symmetrical Bell state of two photons entangled in polarization: $|\Phi^+\rangle = (1/\sqrt{2})\{|x_A,x_B\rangle + |y_A,y_B\rangle\}$ (or, as D.Klyshko named it, a *biphoton*). It is measured in two different stations, where polarization analyzers are set at angles α and β. This means that $|\Phi^+\rangle$ is projected into the *bra* state:

$$\langle\alpha_A,\beta_B| \equiv [cos(\alpha)\langle x_A| + sin(\alpha)\langle y_A|] \otimes$$
$$\otimes [cos(\beta)\langle x_B| + sin(\beta)\langle y_B|] \quad (1)$$

then:

$$\langle\alpha_A,\beta_B|\Phi^+\rangle = (1/\sqrt{2})[cos(\alpha).cos(\beta) + sin(\alpha).sin(\beta)] \quad (2)$$

which is the sum of the probability amplitudes of being detected as a pair of two *x*-photons or of two *y*-photons. The square modulus of this amplitude probability is the observable probability of coincidence: $P^{++} = \frac{1}{2} cos^2(\alpha-\beta)$. This expression violates the Bell's inequalities, and the state is recognized as entangled.

Let assume now that the observer knows if the pair is $|x_A,x_B\rangle$ or $|y_A,y_B\rangle$, thanks to the information available in some auxiliary system H. The state is now:

$$|\Phi^{'}\rangle \equiv (1/\sqrt{2})\{|x_A,x_B\rangle\otimes|1\rangle_H + |y_A,y_B\rangle\otimes|0\rangle_H\} \quad (3)$$

There is no ambiguity in the information stored in H, i.e., $\langle 1|0\rangle_H = 0$. Note that $|\Phi^{'}\rangle$ is a GHZ state. The

probability of coincidence is now:

$$P^{++} = \tfrac{1}{2} \{cos^2(\alpha) \cdot cos^2(\beta) + sin^2(\alpha) \cdot sin^2(\beta)\} \quad (4)$$

which is the sum of the probabilities for the biphoton to be detected as $|x_A,x_B\rangle$ or as $|y_A,y_B\rangle$. The eq.4 does not violate the Bell's inequalities. It corresponds to a mixture of classical (separable) states. Now consider the basis of H with well-defined parity:

$$|+\rangle_H \equiv (1/\sqrt{2})[|1\rangle_H + |0\rangle_H] \,,\, |-\rangle_H \equiv (1/\sqrt{2})[|1\rangle_H - |0\rangle_H] \quad (5)$$

it is immediate that:

$$|\Phi'\rangle = (1/\sqrt{2})\{|\Phi^+\rangle \otimes |+\rangle_H + |\Phi^-\rangle \otimes |-\rangle_H\} \quad (6)$$

where $|\Phi^-\rangle = (1/\sqrt{2})\{|x_A,x_B\rangle - |y_A,y_B\rangle\}$ is the biphoton with parity opposite to $|\Phi^+\rangle$. If a measurement on the basis $\{|+\rangle_H, |-\rangle_H\}$ is performed on H, the information on the pair's polarization is erased. According to the result obtained in this measurement, the data corresponding to eq.4 can be separated in two sets that do show entanglement. This is easily understood by noting that the eq.4 can also be written:

$$P^{++} = \tfrac{1}{2} \{\tfrac{1}{2} cos^2(\alpha-\beta) + \tfrac{1}{2} cos^2(\alpha+\beta)\} \quad (7)$$

where the first (second) term into the keys is the coincidence probability corresponding to the state $|\Phi^+\rangle$ ($|\Phi^-\rangle$). This scheme is hence the precise analogous, for a biphoton, to the quantum eraser for a single photon. The entangled-separable properties are in the same relationship than the wave-particles ones. One or the other is observed depending on the information available on the system.

## 3. Proposal of implementation using three-photon spontaneous decay.

In degenerate non-collinear three-photon spontaneous decay (3psd), one photon of wavelength $\lambda_p$ decays into three photons of wavelength $3\times\lambda_p$, which leave the crystal at directions determined by the phase-matching condition [7]. The 3psd process has been generally disregarded because the involved nonlinearity (which is proportional to the third-order susceptibility) is small, and hence the generated fluorescence is faint. Yet, the rate of heralded biphotons produced by 3psd is calculated to be higher than what can be obtained from multiple 2psd standard processes. Besides, the generated state has no fundamental limitation on its purity [7,8]. A practical reason for not using 3psd was that Silicon single photon detectors are efficient around 700 nm. Therefore, they require pumping the crystal with UV radiation in the 200-300 nm region. In early tests using the fourth harmonic of a Nd:YVO$_4$ pump laser (266 nm), the resulting strong fluorescence of impurities in the crystal masked the 3psd signal. Nowadays, InGaAs/InP single photon detectors efficient in the 1200 nm region are available. Pumping with laser diodes at 405 nm (a pump wavelength where the impurities' fluorescence is known to be tolerable) becomes now feasible.

The convenient material for 3psd is a centro-symmetrical crystal. In such a crystal the second-order susceptibility tensor vanishes and, in consequence, also the 2psd fluorescence, which may otherwise mess with the radiation of interest here. In what follows, I assume that a pump beam at 405 nm, polarized parallel to the extraordinary axis, decays in Calcite into two ordinary plus one extraordinary beam at 1215 nm: $e\rightarrow(o,o',e)$. This process is supposed to have the highest value of the third-order susceptibility tensor in Calcite in the optical region [9]. The phase-matching angle $\Psi$ is hence calculated from the expressions:

$$3\times n_e(405\text{ nm}) = n_e(1215\text{ nm}) + 2\times n_o(1215\text{ nm}) \quad (8)$$

$$1/n_e^2 = cos^2(\Psi)/n_o^2 + sin^2(\Psi)/n_e^2 \quad (9)$$

where $n_o$, $n_e$ are the indexes of refraction for the ordinary and extraordinary beams at the indicated wavelengths, and $\Psi$ is the angle between the vector of propagation of the pump beam and the optical axis. The solution is $\Psi_{pm} = 31.8°$. The data and methods to calculate the phase matching condition, the distribution of the 3psd fluorescence and to estimate the amount of generated radiation are detailed in [7].

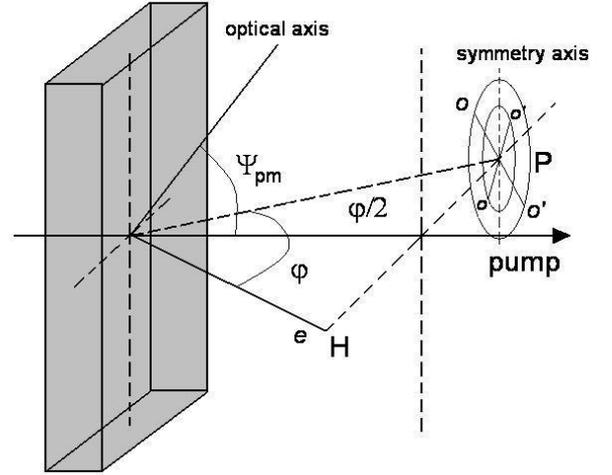

Figure 1: Scheme of non-collinear degenerate $e\rightarrow(o,o',e)$ 3psd. The (e-polarized) pump beam of wavelength $\lambda_p$ crosses the crystal at the phase-matching angle $\Psi_{pm}$ with the optical axis. If an e-photon of wavelength $3\times\lambda_p$ is detected at an angle $\varphi$, then two o-photons $(o,o')$ of wavelength $3\times\lambda_p$ are available at each side of the indicated symmetry axis. This axis is placed at an angle $\tfrac{1}{2}\varphi$ on the other side of the plane determined by the optical axis and the pump beam. The $(o,o')$ photons are placed on the opposite sides of circles concentric to the point P, in the same way that in Type-I 2psd with the pump beam centered in P.

The phase-matching condition for paraxial angles determines that, if one of the 3psd photons is detected at an angle $\varphi$ from the direction of the pump beam, then the other two are found on concentric rings, similar to the one observed in 2psd, but with the center in a point placed at an angle $\tfrac{1}{2}\varphi$ on the opposite side of the pump beam (see the Figure 1). The ring structure is the one of 2psd fluorescence of Type-I (Type II) phase matching if the two photons are the *o,o'* pair (the *o,e* pair). Therefore, if a trigger or herald photon polarized as *e* is detected at the point H, then pairs of *o*-polarized photons are detected on the opposite sides of circles centered in a point P. The point P plays here the same role than the position of the pump beam in 2psd. The efficiency of collection of those pairs at each side of the symmetry axis determines the reliability of the herald. Note that in 2psd the detection of a "signal" photon requires the careful alignment of the other detector to collect the corresponding "idler" photon. In 3psd instead, the detection of the herald determines the position of the symmetry axis, and then *all* the photons at each side of this axis must be collected.

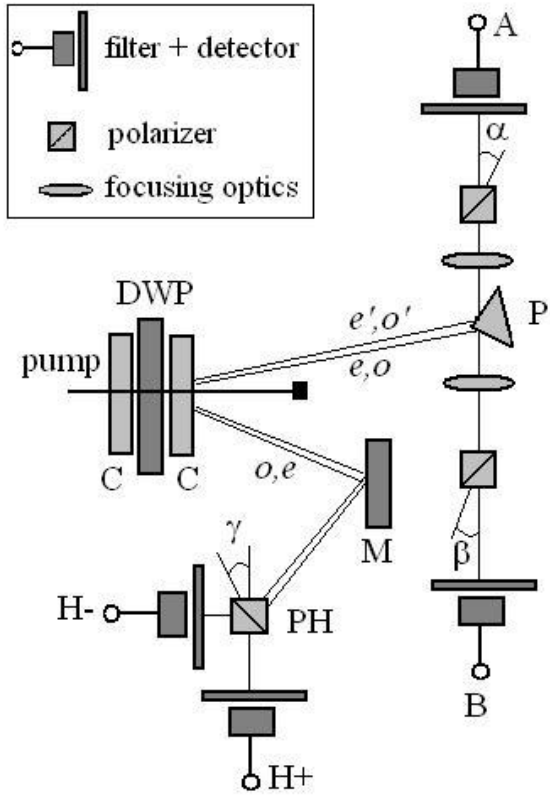

Figure 2: Sketch of the proposed setup. C: calcite crystals; DWP: dichroic waveplate (see Section 3); P: prism placed with the vertex at the symmetry axis of the Fig.1; the interferential filters have maximal transmission at $3\times\lambda_p$.

The disentanglement eraser scheme is now at hand (see the Figure 2). Two identical calcite crystals are placed with their axes parallel to each other, with a dichroic waveplate, similar to the ones used for third harmonic generation, between them. The waveplate is half wavelength for $3\times\lambda_p$ and full wavelength for $\lambda_p$, and its axis is oriented at 45° of the crystals' axes. The purpose of the waveplate is to flip the polarization of the pair produced in the first crystal, from (say) (*e=y*, *o=x*) to (*o=y*, *e=x*) without affecting the phase-matching condition in the second crystal. In this way, the three beams leaving the "sandwich" of crystals have undefined polarization, but they are entangled as in the eq.3.

The rate of triple coincidences {A, B, H} is now recorded as a function of the analyzers' angles α, β. If the herald photon is detected directly, by placing the H detector before the polarizer PH, then the probability of a coincidence at A,B as a function of (α, β) is given by the eq.4. The photons detected at A,B are hence not entangled. Suppose now that the herald photon passes through the polarizer PH and is detected at H+ (transmitted port) or H- (reflected port). If PH is oriented such that a signal in H+ corresponds to the detection of a *y*-polarized photon (i.e., γ=90° from the *x* axis), then the probability of coincidence at A,B is given by the first term in the rhs of eq.4. A detection at H- heralds coincidences at A,B given by the second term in the rhs of eq.4. None of these sets show entanglement.

If γ=45° instead, then a detection at H+ (H-) heralds coincidences at A,B whose probability is given by the first (second) term in the rhs of eq.7, and are hence entangled (state |Φ+⟩ or |Φ-⟩). The result is not affected by the time ordering the detections. It is the *information* erased during the appropriate measurement of the herald what allows the observer to separate an otherwise non-entangled set of data into two sets corresponding to two entangled states with opposite parity.

To the best of my knowledge, this is the simplest proposal to implement the disentanglement eraser.

### 4. Some distinctive features of the proposed setup.

For an arbitrary value of γ, a detection at H+ heralds the state $sin(\gamma)\times|x_A,x_B\rangle + cos(\gamma)\times|y_A,y_B\rangle$, and at H-, the state $cos(\gamma)\times|x_A,x_B\rangle - sin(\gamma)\times|y_A,y_B\rangle$. The Concurrence [10] of these states is $|sin(2\gamma)|$. Therefore, the state at A,B is entangled only if the herald is measured at certain values of γ. These values correspond to measurements that completely erase the information carried by the herald. Nevertheless, the complete erasure of this information is not enough to retrieve *observed* entanglement, as it is discussed below.

The scheme allows the preparation of different states at A,B by appropriately choosing the measurement performed on the herald. This is easily achieved in this full-optical setup. For example, by inserting a quarter waveplate before PH, the herald is projected into the circular polarization basis {|R⟩,|L⟩} and the state at A,B becomes:

$$|\Phi'\rangle = (1/\sqrt{2})\{|\xi^+\rangle\otimes|R\rangle_H + |\xi^-\rangle\otimes|L\rangle_H\} \quad (10)$$

where:

$$|\xi^\pm\rangle = (1/\sqrt{2})\{|x_A,x_B\rangle \pm i\,|y_A,y_B\rangle\} \quad (11)$$

In this version of the setup, the information on the linear polarization of the herald is completely erased and the state at A,B is maximally entangled, the same as it was when γ=45°. However, the detection of |R⟩ or |L⟩ does *not* allow separating the data obtained in A,B

into two sets showing entanglement. In fact, the coincidence probabilities produced by $|\xi^\pm\rangle$ are given by eq.4, i.e., they correspond to a mixture of classical states, not violating Bell's inequalities and showing no sign of entanglement. In order to verify that the state at A,B is in fact entangled, its tomography reconstruction should be performed. A simpler alternative is to insert a quarter waveplate before the analyzer in A *or* in B. The coincidence data can now be separated (according if the herald is observed to be $|R\rangle$ or $|L\rangle$) into two sets that do violate the Bell's inequalities, as before.

Some practical details: note that, due to geometry, the pairs emitted in the first crystal are laterally displaced from the pairs emitted in the second crystal. Besides, if they are *o*-photons in the first crystal, they become *e*-photons in the second, and hence suffer an additional displacement due to walk-off. To keep the emissions from the first and the second crystal indistinguishable, restrictions analogous to the ones for 2psd in two crossed Type-I crystals [11] must be taken into account. In short: the pump beam diameter *d* must hold to the condition $d > \varphi \times (l_c + l_w)$, where $l_c$ is the crystals' and $l_w$ the wave-plate's length, and $\varphi$ is the angle at which the herald photon is detected (see Fig.1). In the practice, a larger value of *d* means a more extended source and hence a smaller attainable efficiency of collection of the heralded pairs, so it must not be increased beyond what is strictly necessary. Be also aware that $\varphi$ cannot be made too small. Otherwise, part of the heralded radiation would mess with the bright pump beam, then becoming difficult to detect. Also, be warned that the characteristics and positions of the prism P and the mirror M in the Fig.2 are for the drawing's convenience. If these devices were used at the drawn angles, they would polarize the beams and spoil the GHZ state $|\Phi'\rangle$. Care must be taken to aim the beams using optical surfaces at small incidence angles, in order to avoid undesired polarization by reflection.

The described setup uses a "sandwich" of three elements, but a single crystal may suffice. This can be done by pumping with a polarization at an intermediate angle (between *o* and *e*), and taking advantage of the interference between the processes: $e \rightarrow (o, o', e)$ and $o \rightarrow (e, e', o)$. Adjusting the intermediate angle allows the compensation of the different values of the third-order susceptibility tensor, so that the amplitude probabilities of both processes are equal. The problem is that the phase matching angle in Calcite is different for each of the two processes. The emission angles of the 3psd fluorescence are different, and hence the processes are distinguishable (and do not interfere). Nevertheless, using a different crystal and/or temperature tuning to equalize the phase matching angles may provide a way to implement the disentanglement eraser with a single optical element.

In summary: a proposal for the full optical implementation of the disentanglement eraser scheme using photons generated by 3psd is described. It is simpler to achieve than the reported alternatives. It is feasible thanks to the relatively recent availability of efficient single-photon detectors at 1200 nm. The scheme demonstrates the physical importance of information in quantum systems, and is of general interest in the fields of quantum information and quantum computation.

**Acknowledgements.**

Many thanks to Gustavo Bosyk and Federico Holik (Universidad Nacional de La Plata, Argentina) for their review of the first version of this manuscript, and their useful suggestions and advices. This contribution received support from the contract PIP2011-077 "Desarrollo de láseres sólidos bombeados por diodos y de algunas de sus aplicaciones", CONICET (Argentina).

**References.**